\documentclass[superscriptaddress,showpacs, 12pt]{revtex4}
\usepackage{amsfonts}
\usepackage{graphicx, amsmath}
\usepackage{longtable}
\makeatletter

\newcommand{\Rmnum}[1]{\expandafter\@slowromancap\romannumeral #1@}
\makeatother
\begin{document}
\title[Short Title]{Fast generation of three-dimensional entanglement between two spatially separated atoms via invariant-based
shortcut}
\author{Jin-Lei Wu}
\affiliation{Department of Physics, College of Science, Yanbian
University, Yanji, Jilin 133002, People's Republic of China}
\author{Chong Song}
\affiliation{Department of Physics, College of Science, Yanbian
University, Yanji, Jilin 133002, People's Republic of China}
\author{Xin Ji\footnote{E-mail: jixin@ybu.edu.cn}}
\affiliation{Department of Physics, College of Science, Yanbian
University, Yanji, Jilin 133002, People's Republic of China}
\author{Shou Zhang}
\affiliation{Department of Physics, College of Science, Yanbian
University, Yanji, Jilin 133002, People's Republic of China}

\begin{abstract}
A scheme is proposed for the fast generation of three-dimensional entanglement between two atoms trapped in two cavities connected by a fiber via invariant-based shortcut to adiabatic passage. With the help of quantum Zeno dynamics, the technique of invariant-based shortcut is applied for the generation of two-atom three-dimensional entanglement. The numerical simulation results show that the target state can be generated in a short time with a high fidelity and the scheme is robust against the decoherence caused by the atomic spontaneous emission, photon leakage, and the variations in the parameters. Moreover, the scheme may be possible to be implemented with the current experimental technology.
\end{abstract}
\maketitle
\section{Introduction}
As one of the most interesting features of quantum mechanics, quantum entanglement plays a significant role in quantum mechanics because it not only holds the power for demonstration of the quantum nonlocality against local hidden variable theory~\cite{ABN1935, DMA1990}, but also is an important part of quantum information processing and quantum computing, such as quantum cryptography~\cite{A1991}, quantum teleportation~\cite{CGC1993}, quantum dense coding~\cite{CS1992}, and so on.

Recently, high-dimensional entanglement is becoming more and more important due to their superior security than qubit systems, especially in the aspect of quantum key distribution. Besides, it has been demonstrated that violations of local realism by two entangled high-dimensional systems are stronger than that by two-dimensional systems~\cite{DPMW2000}. So a lot of efforts have been done in theory and experiment for generating high-dimensional entanglement via different techniques
~\cite{WG2011,WAL2011,LP2011,LP2012,QCW2013,SJR2013,XQ201405,XQ201402,SL2014,YLS2015,AAGA2001,AGA2002}. For instance, Li $et~al.$ implemented the two-atom three-dimensional entanglement by quantum Zeno dynamics~(QZD) in 2011~\cite{WG2011,WAL2011}, Chen $et~al.$ prepared the two-atom three-dimensional entanglement using stimulated Raman adiabatic passage~(STIRAP) in 2011 and 2012~\cite{LP2011,LP2012}, Su $et~al.$ generated the two-atom three-dimensional entanglement via atomic spontaneous emission and cavity decay in 2014~\cite{SL2014}, and Vaziri $et~al.$ experimentally implemented two-photon three-dimensional entanglement for quantum communication in 2002~\cite{AGA2002}. Among these techniques, there are two famous techniques for their robustness against decoherence in proper conditions. One is STIRAP~\cite{LP2011,LP2012,YLS2015}, and the other is QZD~\cite{WG2011,WAL2011,SJR2013}. STIRAP is widely used in time-dependent interacting field because of the robustness against the atomic spontaneous emission and variations in the experimental parameters. But it usually requires a relatively long interaction time, so the decoherence would destroy the intended dynamics and finally lead to an error result. Different from the adiabatic passage, QZD is usually robust against photon leakage but sensitive to the atomic spontaneous emission and variations in the experimental parameters. Thus some of the researchers introduce detuning between the atomic transition to restrain the influence of atomic spontaneous emission~\cite{SJR2013}. However, that also increases the operation time. Therefore, reducing the time of dynamics towards the perfect final outcome is necessary and perhaps the most effective method to essentially fight against the dissipation caused by noise or losses accumulated during the operational processes. In order to solve this problem, recently researchers pay more attention to ``shortcut to adiabatic passage (STAP)'' which employs a set of techniques to speed up a slow quantum adiabatic process~\cite{XASA2010,AC2013,MYLJ2014,YHC2014,YYQ2014,YLQ2015,YLC2015,YLX2015,JYC2016,YHC2016}, in which Chen $et~al.$ proposed the shortcut to adiabatic passage in two- and three-level atoms in 2010~\cite{XASA2010}, then Chen $et~al.$ implemented fast population transfer and entangled states' preparation and transition in multiparticle systems via shortcut to adiabatic passage in 2014 ~\cite{YHC2014,YYQ2014}, and Lin $et~al.$ generated the two-atom three-dimensional via invariants-based shortcut in 2016~\cite{JYC2016}.

In this paper, based on the Lewis-Riesenfeld invariants and QZD we construct an effective shortcut to adiabatic passage for fast generating three-dimensional entanglement between two atoms trapped in two spatially separated cavities connected by a fiber. The generation of two-atom three-dimensional entanglement in our scheme is implemented within a short time and the strict numerical simulations demonstrate that our scheme is insensitive to the decoherence caused by the atomic spontaneous emission, photon leakage, and the variations in the parameters. In particular, compared with previous work using the same technique which prepared two-atom three-dimensional entanglement by two steps using a superposition state as the initial state ~\cite{JYC2016}, our scheme is more feasible because of easier initial state and one-step implementation of the target state.

This paper is structured as follows. In Section~\ref{sec2}, we give a brief review of QZD and Lewis-Riesenfeld invariants. In Section~\ref{sec3}, we will describe the physical model and the generation of two-atom three-dimensional entanglement via invariant-based shortcut. In Section~\ref{sec4}, we give the numerical simulations and discussion of feasibility of the fast generation of two-atom three-dimensional entanglement in our scheme. Finally, the conclusion is given in Section~\ref{sec5}.

\section{The brief review of quantum Zeno dynamics and Lewis-Riesenfeld invariants}
\label{sec2}

\subsection{Quantum Zeno dynamics}
For the sake of clearness, we first briefly give a review of the quantum Zeno dynamics. Assume that a quantum system's dynamics evolution is governed by the Hamiltonian
\begin{eqnarray}\label{e1}
H_K=H_{\rm obs}+K H_{\rm meas},
\end{eqnarray}
where $H_{\rm obs}$ can be viewed as the Hamiltonian of the quantum system investigated and $H_{\rm meas}$ as an additional interaction Hamiltonian performing the measurement. $K$ is a coupling constant, and in the strong coupling limit $K\rightarrow \infty$, the whole system is governed by the evolution operator~\cite{PGS2009}
\begin{eqnarray}\label{e2}
U(t)=\lim_{K\rightarrow \infty}{\rm exp}[-it\sum_{n}(K\lambda_nP_n+P_nH_{\rm obs}P_n)],
\end{eqnarray}
where $\sum_{n}P_nH_{obs}P_n$ is Zeno Hamiltonian, $P_n$ is one of the eigenprojections of $H_{\rm meas}$ with
eigenvalues $\lambda_n$($H_{\rm meas} = \sum_{n}\lambda_nP_n$). Interestingly, it is easy to deduce~\cite{PS2002, PGS2009} that the system state will remain in the same Zeno subspace as that of its initial state. Specially, if the system is initially in the dark state $|\Psi_d\rangle$ of $H_{\rm meas}$, i.e., $H_{\rm meas}|\Psi_d\rangle=0$, the evolution operator reduces to~\cite{XLS2010}
\begin{eqnarray}\label{e3}
U(t)=\lim_{K\rightarrow \infty}{\rm exp}(-itP_nH_{\rm obs}P_n).
\end{eqnarray}

\subsection{Lewis-Riesenfeld invariants}

Here we give a brief description about Lewis-Riesenfeld invariants theory~\cite{HRL1969}. A quantum system is governed by a time-dependent Hamiltonian $H(t)$, and the corresponding time-dependent Hermitian invariant $I(t)$ satisfies
\begin{eqnarray}\label{4}
i\hbar\frac{\partial I(t)}{\partial t}&=&[H(t),I(t)].
\end{eqnarray}
The solution of the time-dependent Schr\"odinger equation $i\hbar\frac{\partial|\Psi(t)\rangle}{\partial t}=H(t)|\Psi(t)\rangle$ can be expressed by a superposition of invariant $I(t)$ dynamical modes $|\Phi_{n}(t)\rangle$
\begin{eqnarray}\label{5}
|\Psi(t)\rangle&=&\sum_n C_n e^{i\alpha_n}|\Phi_{n}(t)\rangle,
\end{eqnarray}
where $C_n$ is time-independent amplitude, $\alpha_n$ is the Lewis-Riesenfeld phase, and $|\Phi_{n}(t)\rangle$ is one of the
orthogonal eigenvectors of the invariant $I(t)$ satisfying $I(t)|\Phi_{n}(t)\rangle=\lambda_n|\Phi_{n}(t)\rangle$ with a real eigenvalue $\lambda_n$. The Lewis-Riesenfeld phases
are defined as
\begin{eqnarray}\label{6}
\alpha_n(t)&=&\frac{1}{\hbar}\int_0^t dt^\prime
\langle\Phi_n(t^\prime )|i\hbar\frac{\partial}{\partial t^\prime
}-H(t^\prime )|\Phi_n(t^\prime )\rangle.
\end{eqnarray}

\section{The physical model and generation of two-atom three-dimensional entanglement via invariant-based shortcut}
\label{sec3}

\begin{figure}[htb]\centering
  \includegraphics[scale=0.5]{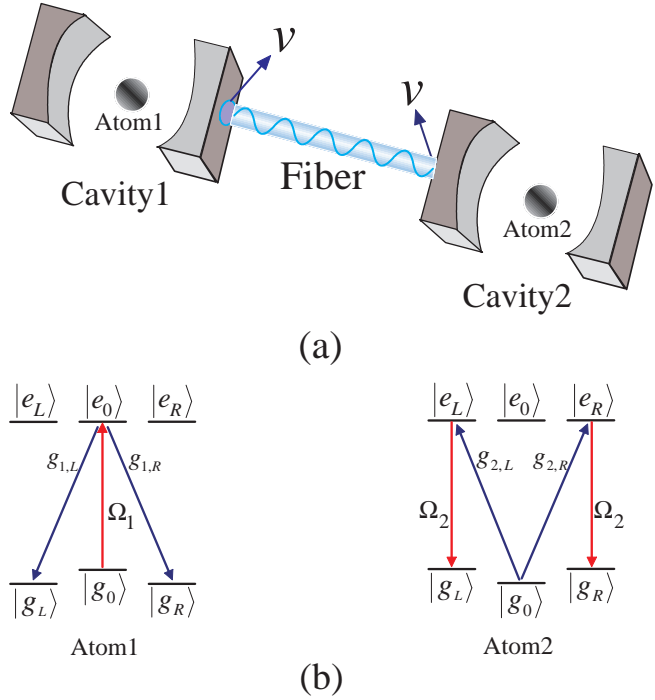}\\
  \caption{(Color online) \textbf{(a)}~The schematic setup for generating two-atom three-dimensional entanglement. \textbf{(b)}~The level configurations and relevant transitions of two atoms. }\label{f1}
\end{figure}
The schematic setup for generating two-atom three-dimensional entanglement is shown in Fig.~\ref{f1}\textbf{(a)}. Two atoms are trapped in two spatially separated optical cavities connected by a fiber. Under the short fiber limit $(lv)/(2\pi c)\leq 1$, only the resonant mode of the fiber interacts with the cavity mode~\cite{SMB2006}, where $l$ is the length of the fiber and $v$ is the decay rate of the cavity field into a continuum of fiber modes. The level configurations and relevant transitions of two atoms are shown in Fig.~\ref{f1}\textbf{(b)}. The tripod-type atom1 with two degenerate ground states $|g_L\rangle$ and $|g_R\rangle$ and the $M$-type atom2 with two degenerate excited states $|e_L\rangle$ and $|e_R\rangle$ are trapped in two double-mode cavities, respectively. The atomic transitions $|e_0\rangle_1\leftrightarrow|g_L\rangle_1~(|g_R\rangle_1)$ and $|g_0\rangle_2\leftrightarrow|e_L\rangle_2~(|e_R\rangle_2)$ are coupled resonantly to the left-circularly~(right-circularly) polarized modes of cavity1 and cavity2 with corresponding coupling constants $g_{1,L}~(g_{1,R})$ and $g_{2,L}~(g_{2,R})$, respectively. The transitions $|g_0\rangle_1\leftrightarrow|e_0\rangle_1$ and $|e_{L(R)}\rangle_2\leftrightarrow|g_{L(R)}\rangle_2$ are driven resonantly by classical laser fields with the time-dependent Rabi frequencies $\Omega_1(t)$ and $\Omega_2(t)$, respectively.

Then, in the interaction picture, the interaction Hamiltonian of the whole system can be written as~(assuming $\hbar=1$ for simplicity)~\cite{WAL2011,LP2012}:
\begin{eqnarray}\label{e7}
H_{\rm total}&=&H_{al}+H_{acf},\nonumber\\
H_{al}&=&\Omega_1(t)|e_0\rangle_1\langle{g_0}|+\Omega_2(t)(|e_L\rangle_2\langle{g_L}|+|e_R\rangle_2\langle{g_R}|)+\rm H.c.,\nonumber\\
H_{acf}&=&\sum_{i=L,R}\Big[g_{1,i}a_{1,i}|e_0\rangle_{1}\langle{g_i}|+g_{2,i}a_{2,i}|e_i\rangle_{2}\langle{g_0}|+v b_i(a^{\dag}_{1,i}+a^{\dag}_{2,i})\Big]
+\rm H.c.,
\end{eqnarray}
where $H_{\rm total}$ is the total Hamiltonian of the whole system, $H_{al}~(H_{acf})$ is the interaction between the atoms and the classical laser fields~(the cavity-fiber system), $v$ is the coupling strength between the cavity modes and the fiber modes, $b_{L~(R)}$ is the annihilation operator of left-circularly~(right-circularly) polarized mode of the fiber, and $a_{1,L~(R)}~(a^{\dag}_{1,L~(R)})$ is the annihilation operator of left-circularly~(right-circularly) polarized mode of cavity1~(cavity2). For simplicity, we assume $g_{1,L~(R)}$ and $g_{2,L~(R)}$ are real, and $g_{1,L~(R)}=g_{2,L~(R)}=g$.

Suppose that the total system is initially in the state $|\phi_1\rangle=|g_0\rangle_1|g_0\rangle_2|0\rangle_1|0\rangle_f|0\rangle_2$ denoting that two atoms are in the states $|g_0\rangle_1$ and $|g_0\rangle_2$ respectively, and the two cavities and the fiber all in the vacuum state. Afterwards, governed by the total Hamiltonian in Eq.~(\ref{e7}), the whole system evolves in the subspace spanned by
\begin{eqnarray}\label{e8}
    |\phi_1\rangle&=&|g_0\rangle_1|g_0\rangle_2|0\rangle_1|0\rangle_f|0\rangle_2, ~\nonumber
    |\phi_2\rangle=|e_0\rangle_1|g_0\rangle_2|0\rangle_1|0\rangle_f|0\rangle_2,\nonumber\\
    |\phi_3\rangle&=&|g_L\rangle_1|g_0\rangle_2|L\rangle_1|0\rangle_f|0\rangle_2, ~\nonumber
    |\phi_4\rangle=|g_R\rangle_1|g_0\rangle_2|R\rangle_1|0\rangle_f|0\rangle_2,\nonumber\\
    |\phi_5\rangle&=&|g_L\rangle_1|g_0\rangle_2|0\rangle_1|L\rangle_f|0\rangle_2, ~\nonumber
    |\phi_6\rangle=|g_R\rangle_1|g_0\rangle_2|0\rangle_1|R\rangle_f|0\rangle_2,\nonumber\\
    |\phi_7\rangle&=&|g_L\rangle_1|g_0\rangle_2|0\rangle_1|0\rangle_f|L\rangle_2, ~\nonumber
    |\phi_8\rangle=|g_R\rangle_1|g_0\rangle_2|0\rangle_1|0\rangle_f|R\rangle_2,\nonumber\\
    |\phi_9\rangle&=&|g_L\rangle_1|e_L\rangle_2|0\rangle_1|0\rangle_f|0\rangle_2, ~\nonumber
    |\phi_{10}\rangle=|g_R\rangle_1|e_R\rangle_2|0\rangle_1|0\rangle_f|0\rangle_2,\nonumber\\
    |\phi_{11}\rangle&=&|g_L\rangle_1|g_L\rangle_2|0\rangle_1|0\rangle_f|0\rangle_2, ~
    |\phi_{12}\rangle=|g_R\rangle_1|g_R\rangle_2|0\rangle_1|0\rangle_f|0\rangle_2.
\end{eqnarray}
Obviously, the system is initially in the dark state of $H_{acf}$, i.e., $H_{acf}|\phi_1\rangle=0$. Therefore, under the limit condition $\Omega_1(t)$,~$\Omega_2(t)\ll g$ and by means of the technique of QZD, the whole system can approximatively evolve in an invariant Zeno subspace consisting of dark states corresponding to the zero eigenvalue of $H_{acf}$:
 \begin{eqnarray}\label{e9}
    H_P=\{|\phi_1\rangle, |\Psi_D\rangle, |\phi_{11}\rangle, |\phi_{12}\rangle \},
\end{eqnarray}
corresponding to the projections
\begin{equation}\label{e10}
    P^\alpha=|\alpha\rangle\langle\alpha|,~~~~(|\alpha\rangle\in H_{P}).
\end{equation}
Here,
\begin{eqnarray}\label{e11}
    |\Psi_D\rangle&=&\frac{1}{\sqrt{3v^2+2g^2}}[v|\phi_2\rangle-g(|\phi_5\rangle+|\phi_6\rangle)+v(|\phi_9\rangle+|\phi_{10}\rangle)].
\end{eqnarray}
Therefore, the system Hamiltonian can be rewritten as the following form based on Eq.~(\ref{e3}):
\begin{eqnarray}\label{e12}
H_{\rm total}&\simeq&\sum_{\alpha}P^\alpha H_{al}P^\alpha \nonumber\\
&=&\frac{v}{\sqrt{3v^2+2g^2}}[\Omega_1(t)|\phi_1\rangle\langle\Psi_D|+\Omega_2(t)(|\phi_{11}\rangle+|\phi_{12}\rangle)\langle\Psi_D|]+\rm H.c..
\end{eqnarray}
 Here setting $v=g$,  we can obtain an effective Hamiltonian of the system
\begin{eqnarray}\label{e13}
H_0(t)=\frac{1}{\sqrt5}(\Omega_1(t)|\Psi_1\rangle+\Omega'_2(t)|\Psi_2\rangle)\langle\Psi_D|+\rm H.c..
\end{eqnarray}
in which $|\Psi_1\rangle=|\phi_1\rangle$, $|\Psi_2\rangle=\frac{1}{\sqrt2}(|\phi_{11}\rangle+|\phi_{12}\rangle)$, and $\Omega'_2(t)=\sqrt2~\Omega_2(t)$. The target state we expect is the state $|\Psi_{3D}\rangle=\frac{1}{\sqrt3}|\Psi_1\rangle+\frac{\sqrt2}{\sqrt3}|\Psi_2\rangle=\frac{1}{\sqrt3}(|g_0\rangle_1|g_0\rangle_2-|g_L\rangle_1|g_L\rangle_2-|g_R\rangle_1|g_R\rangle_2)$.

In order to construct the invariant-based shortcut for generating three-dimensional entanglement, we need to find out the Hermitian invariant operator $I(t)$, which satisfies $i\hbar \frac{\partial I(t)}{\partial t}=[H_{0}(t),I(t)]$. Since $H_{0}(t)$ possesses SU(2) dynamical symmetry, $I(t)$ can be easily given by~\cite{MA2001,XCE2011}
\begin{eqnarray}\label{e16}
I(t)&=&=\frac{1}{\sqrt5}\chi\left(
\begin{array}{ccc}
0                                        & \cos\nu\sin\beta                   & -i\sin\nu    \\
\cos\nu\sin\beta                         & 0                                  & \cos\nu\cos\beta \\
i\sin\nu                                 & \cos\nu\cos\beta                   & 0
\end{array}
\right).
\end{eqnarray}
where $\chi$ is an arbitrary constant with units of frequency to keep $I(t)$ with dimensions of energy, $\nu$ and $\beta$ are time-dependent auxiliary parameters which satisfy the equations
\begin{eqnarray}\label{e17}
\dot{\nu}&=&\frac{1}{\sqrt5}(\Omega_{1}(t)\cos\beta-\Omega'_{2}(t)\sin\beta),\nonumber\\
\dot{\beta}&=&\frac{1}{\sqrt5}\tan\nu(\Omega'_{2}(t)\cos\beta+\Omega_{1}(t)\sin\beta).
\end{eqnarray}
Then we can deduce $\Omega_{1}(t)$ and $\Omega'_{2}(t)$ easily as follows:
\begin{eqnarray}\label{e18}
\Omega_{1}(t)&=&\sqrt5(\dot{\beta}\cot\nu\sin\beta+\dot{\nu}\cos\beta),\nonumber\\
\Omega'_{2}(t)&=&\sqrt5(\dot{\beta}\cot\nu\cos\beta-\dot{\nu}\sin\beta).
\end{eqnarray}
The solution of Shr\"{o}dinger equation $i\hbar\partial|\Psi(t)\rangle/\partial t=H_{\rm eff}(t)|\Psi(t)\rangle$ with respect to the instantaneous eigenstates of $I(t)$ can be written as $|\Psi(t)\rangle=\sum_{n=0,\pm}C_ne^{i\alpha_n}|\phi_n(t)\rangle$, where $\alpha_n(t)$ is the Lewis-Riesenfeld phase in Eq.~(\ref{6}), $C_n=\langle\phi_n(0)|\Psi_1\rangle$, and $|\phi_n(t)\rangle$ is the eigenstate of the invariant $I(t)$ as
\begin{eqnarray}\label{e19}
|\phi_0(t)\rangle=\left(
\begin{array}{c}
\cos\nu\cos\beta\\
-i\sin\nu\\
-\cos\nu\sin\beta
\end{array}\right), ~~\rm and ~~
|\phi_\pm(t)\rangle=\frac{1}{\sqrt{2}}\left(
\begin{array}{c}
\sin\nu\cos\beta\pm i\sin\beta\\
i\cos\nu\\
-\sin\nu\sin\beta\pm i\cos\beta
\end{array}\right).
\end{eqnarray}
In order to determinate $\nu$ and $\beta$, we impose the boundary conditions to satisfy $[H_0(0), I(0)]=0$ and $[H_0(t_f), I(t_f)]=0$~($t_f$ is the operation time), which give $\Omega_1(0)=0$ and $\Omega_1(t_f)=\sqrt2\Omega'_2(t_f)$. Based on these discussions and to avoid infinite Rabi frequencies,
we set the boundary conditions for $\nu$ and $\beta$ as follows£º
\begin{eqnarray}\label{e20}
\nu(0)=\varepsilon,~\dot{\nu}(0)=0,~\nu(t_f)=\varepsilon,~\dot{\nu}(t_f)=0,~\beta(0)=0,~
\beta(t_f)=\arctan\sqrt2.
\end{eqnarray}
where $\varepsilon$ is a time-independent small value. As a consequence, we can simply choose the parameters as
\begin{eqnarray}\label{e21}
\nu(t)=\varepsilon,~~~~~~~~~\beta(t)=\frac{\arctan\sqrt2t}{t_f},
\end{eqnarray}
providing
\begin{eqnarray}\label{e22}
\Omega_{1}(t)&=&\frac{\sqrt5\arctan\sqrt2}{t_f}\cot\varepsilon\sin\frac{\arctan\sqrt2t}{t_f},\nonumber\\
\Omega'_{2}(t)&=&\frac{\sqrt5\arctan\sqrt2}{t_f}\cot\varepsilon\cos\frac{\arctan\sqrt2t}{t_f}.
\end{eqnarray}

Then we determine the value of $\varepsilon$ by calculating the fidelity
\begin{eqnarray}\label{e23}
F&=&|\langle\Psi_{3D}|\Psi(t_f)\rangle|^2,\nonumber\\
&=&\Big[1-\sin^2\varepsilon\left\{1-\cos(\frac{\arctan\sqrt2}{\sin\varepsilon})\right\}\Big]^2,
\end{eqnarray}
with the Lewis-Riesenfeld phases
\begin{eqnarray}\label{e24}
\alpha_0=0,~~~
\alpha_{\pm}=\mp \frac{\arctan\sqrt2}{\sin\varepsilon}.
\end{eqnarray}
Therefore, for the appropriate Rabi frequencies and the fidelity $F=1$, we can choose
\begin{eqnarray}\label{e25}
\frac{\arctan\sqrt2}{\sin\varepsilon}=2\pi,~~\rm i.e.~~
\varepsilon=\arcsin\left(\frac{\arctan\sqrt2}{2\pi}\right)=0.153.
\end{eqnarray}
Thus, the transformation $|\Psi_1\rangle\rightarrow|\Psi_{3D}\rangle$ is achieved and we have constructed the shortcut to adiabatic passage to speed up the generation of three-dimensional entanglement $|\Psi_{3D}\rangle=\frac{1}{\sqrt3}(|g_0\rangle_1|g_0\rangle_2-|g_L\rangle_1|g_L\rangle_2-|g_R\rangle_1|g_R\rangle_2)$.
\section{Numerical simulations and discussion of feasibility}
\label{sec4}
\begin{figure}\centering
  \includegraphics[scale=0.55]{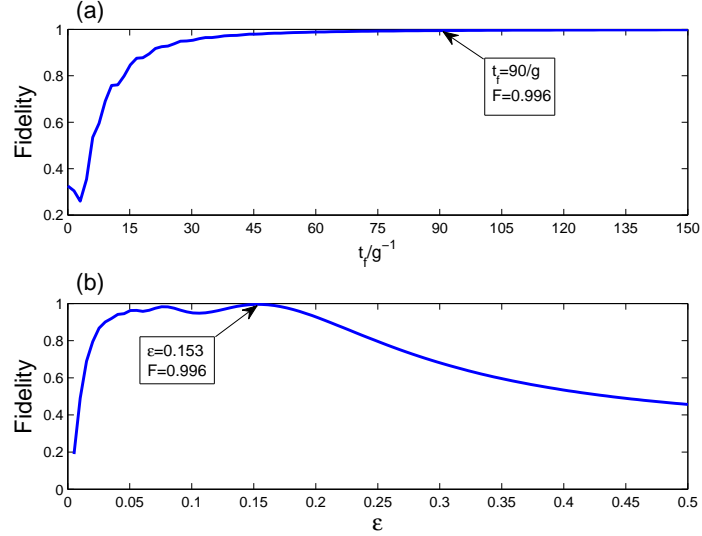}\\
  \caption{\textbf{(a)}~The fidelity of two-atom three-dimensional entanglement versus $t_f/g^{-1}$ via invariant-based shortcut with $\varepsilon=0.153$. \textbf{(b)}~The fidelity of two-atom three-dimensional entanglement versus $\varepsilon$ via invariant-based shortcut with $t_f=90/g$.}\label{f2}
\end{figure}
In the following, we present the numerical simulations of our scheme proposed for generating two-atom three-dimensional entanglement.

In Fig.~\ref{f2}, we plot the fidelity $F=|\langle\Psi_{3D}|\Psi(t_f)\rangle|^2$ versus the operation time $t_f$, where $|\Psi(t_f)\rangle$ is the state of the whole system governed by the total Hamiltonian $H_{\rm total}$ in Eq.~(\ref{e7}) when $t=tf$. From Fig.~\ref{f2}\textbf{(a)} we can see that only in a very short operation time $t_f=90/g$ the fidelity is already almost unity: $F~(t_f=90g^{-1})=0.996$ and from Fig.~\ref{f2}\textbf{(b)} we can find that when $\varepsilon=0.153$ the fidelity is highest. Thus we can choose $t_f=90/g$ and $\varepsilon=0.153$ in the following discussion.
\begin{figure}\centering
  \includegraphics[scale=0.55]{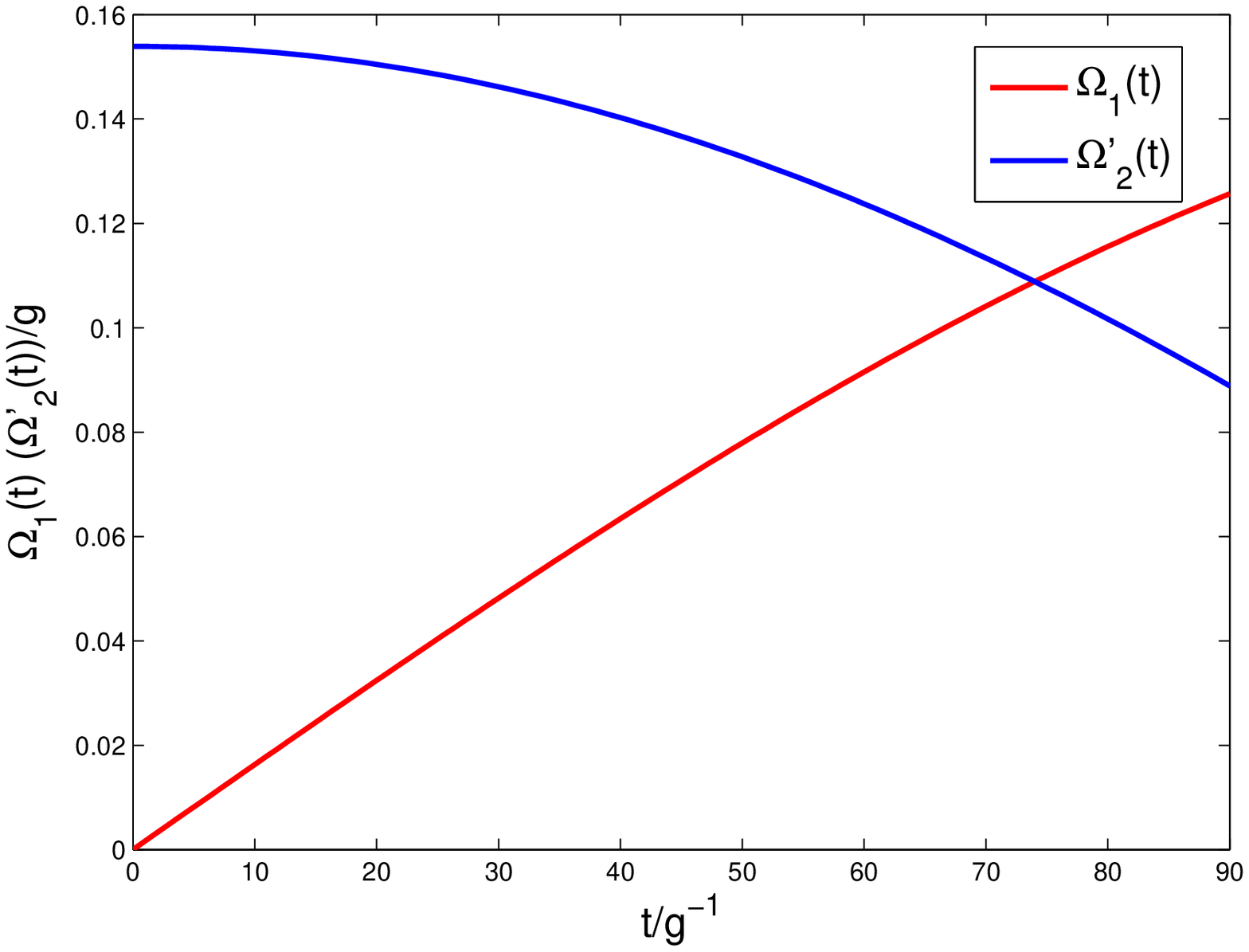}\\
  \caption{(Color online) The time dependence of the the laser fields $\Omega_{1}(t)$~(red line) and $\Omega'_{2}(t)$~(blue line) with the parameters $\varepsilon=0.153$ and $t_f=90/g$.}\label{f3}
\end{figure}
\begin{figure}\centering
  \includegraphics[scale=0.55]{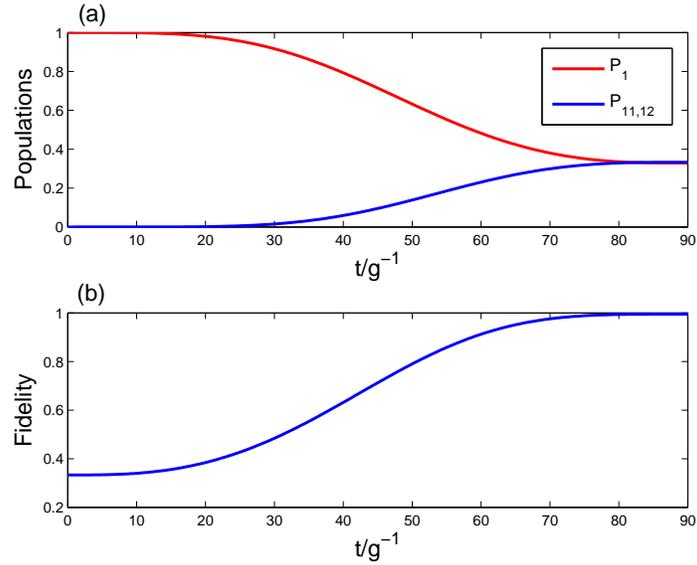}\\
  \caption{(Color online)~\textbf{(a)} The population curves of $|1\rangle$~(red line) and $|11\rangle(|12\rangle)$blue line) and ~\textbf{(b)} the fidelity versus $gt$ with the same parameters as Fig.~\ref{f3}.}\label{f4}
\end{figure}

In Fig.~\ref{f3}, we plot the time dependence of the laser fields $\Omega_{1}(t)$~(red line) and $\Omega'_{2}(t)$~(blue line) based on Eq.~(\ref{e22}). Fig.~\ref{f3} shows that the laser pulses we choose meet the conditions $[H_0(0), I(0)]=0$ and $[H_0(t_f), I(t_f)]=0$ very well, which give $\Omega_1(0)=0$ and $\Omega_1(t_f)=\sqrt2\Omega'_2(t_f)$. Then the population curves and the fidelity versus $gt$ are depicted in Fig.~\ref{f4}~\textbf{(a)} and Fig.~\ref{f4}~\textbf{(b)}, respectively. We can see that, when $t>80g^{-1}$, the population curves of $|\phi_1\rangle$, $|\phi_{11}\rangle$, and $|\phi_{12}\rangle$ coincident reasonably well at $P=\frac{1}{3}$ with each other from Fig.~\ref{f4}~\textbf{(a)} and the fidelity is almost unity from Fig.~\ref{f4}~\textbf{(b)}.

\begin{figure}[htb]\centering
  \includegraphics[scale=0.55]{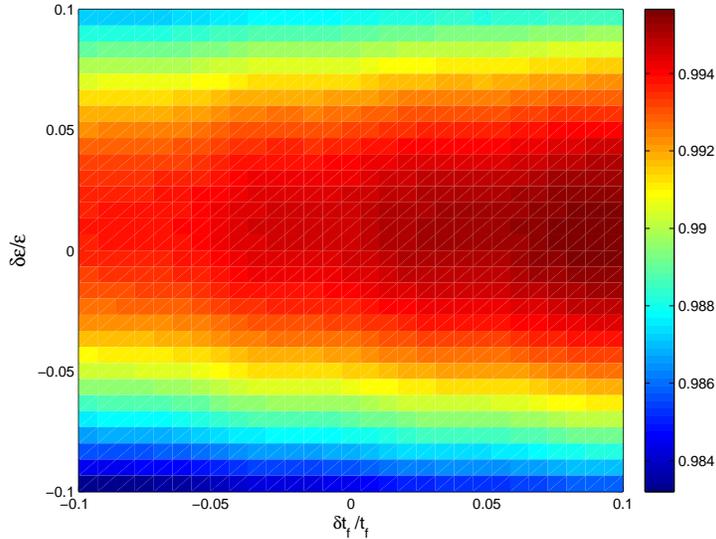}\\
  \caption{(Color online) The fidelity versus $\delta t_f/t_f$ and $\delta\varepsilon/\varepsilon$ with the same parameters as Fig.~\ref{f3}.}\label{f5}
\end{figure}

In the above discussion, the operation and the whole system are perfect and considered as absolutely isolated from the environment and we have omitted the effect of the variations in the parameters and decoherence induced by the atomic spontaneous emissions and photon leakages of the cavities and the fiber. Therefore, for the variations in the parameters, we plot the fidelity versus the variations in $t_f$ and $\varepsilon$ in Fig.~\ref{f5}. Here we define $\delta x= x'-x$ as the deviation of $x$, in which $x$ denotes the ideal value and $x'$ denotes the actual value. In Fig.~\ref{f5}, the fidelity decreases with the increase of $|\delta\varepsilon|$ as described in Fig.~\ref{f2}~\textbf{(b)}. From Eq.~(\ref{e22}), the Rabi frequencies decrease with the increase of the operation time $t_f$. According to the limit condition $\Omega_1$,~$\Omega_2\ll g$ we choose, the values of the Rabi frequencies are the smaller the better, so the operation time $t_f$ is the longer the better as described in Fig.~\ref{f2}~\textbf{(a)}. Therefore, the fidelity increases with the increase of $\delta t_f$ in Fig.~\ref{f5}. Significantly, we notice that the fidelity is over 0.98 even when $\delta t_f/t_f=\delta\varepsilon/\varepsilon=-0.1$ and it shows that our scheme is robust against the variations in the parameters.

Next taking the decoherence induced by the atomic spontaneous emissions and photon leakages of the cavities and the fiber into account, the whole system is dominated by the master equation
\begin{eqnarray}\label{e26}
\dot{\rho}(t)&=&-i[H_{\rm total},\rho(t)]\nonumber\\
\cr&&-\sum_{j=L,R}\frac{\kappa_{j}^{f}}{2}[b_{j}^{\dag}b_{j}\rho(t)-2b_{j}\rho(t)b_{j}^{\dag}+\rho(t)b^\dag b ]\nonumber\\
\cr&&-\sum_{j=L,R}\sum_{i=1,2}\frac{\kappa_{j}^{i}}{2}[a_{ij}^{\dag}a_{ij}\rho(t)-2a_{ij}\rho(t)a_{ij}^{\dag}+\rho(t)a_{ij}^\dag
a_{ij} ]\nonumber\\
\cr&&-\sum_{j=0,L,R}\frac{\gamma_{j}^{1}}{2}[\sigma^{1}_{e_0,e_0}\rho(t)-2\sigma^{1}_{g_j,e_0}\rho(t)\sigma^{1}_{e_0,g_j}+\rho(t)\sigma^{1}_{e_0,e_0}]
\nonumber\\
\cr&&-\sum_{j=0,L,R}\sum_{i=L,R}\frac{\gamma_{j,i}^{2}}{2}[\sigma^{2}_{e_i,e_i}\rho(t)-2\sigma^{2}_{g_j,e_i}\rho(t)\sigma^{2}_{e_i,g_j}+\rho(t)\sigma^{2}_{e_i,e_i}],\nonumber\\
\end{eqnarray}
where $H_{\rm total}$ is the total Hamiltonian in Eq.~(\ref{e7}). $\kappa_{j}^{f}$ is the photon leakage rate of $j$th fiber mode, $\kappa_{j}^{i}$ the photon leakage rate of $j$-circular polarization mode in $i$th cavity, $\gamma_{j}^{1}$ is spontaneous emission rate of the atom1 from the excited state $|e_0\rangle$ to the ground state $|g_j\rangle$, $\gamma_{j,i}^{2}$ is spontaneous emission rate of the atom2 from the excited state $|e_i\rangle$ to the ground state $|g_j\rangle$, $\sigma_{e_j,e_j}=|e_j\rangle\langle e_j|~(j=0,L,R)$, and $\sigma_{e_j,g_j}=|e_j\rangle\langle g_j|$. For simplicity, we assume $\kappa_{j}^{f}=\kappa_{j}^{i}=\kappa$, $\gamma_{j}^{1}=\gamma_{j,i}^{2}=\gamma$.

\begin{figure}[htb]\centering
  \includegraphics[scale=0.55]{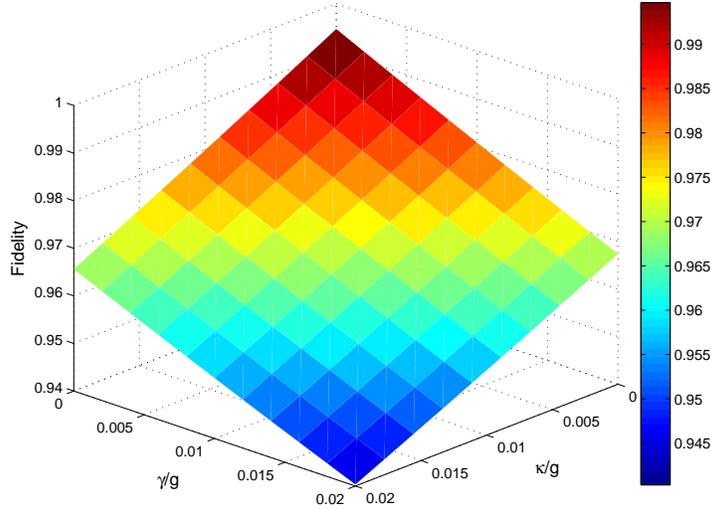}\\
  \caption{(Color online) The fidelity of generation of two-atom three-dimensional entanglement versus $\kappa/g$ and $\gamma/g$.}\label{f6}
\end{figure}

In Fig.~\ref{f6}, we plot the fidelity of generation of two-atom three-dimensional entanglement versus $\kappa/g$ and $\gamma$. As we can see from the decrease of the fidelity with the increases of $\kappa/g$ and $\gamma/g$ in Fig.~\ref{f6}, we know that the influence of atomic spontaneous emissions and that of the cavity-fiber decay on the fidelity are roughly equal and both slight. From Fig.~\ref{f6}, the fidelity is still over 0.94 when $\kappa=\gamma=0.02g$, and thus our scheme is robust against the decoherence induced by the atomic spontaneous emissions and photon leakages of the cavities and the fiber. According to the recent experiments about realizing high-Q cavity and strong atom-cavity coupling~\cite{YAP2011,WZM2010,WZZ2010,SJR2013}, we can choose the experimental parameters as $g/2\pi\sim5.5$~GHz, $\gamma/2\pi\sim4.6$~MHz$\sim0.001g$ and $\kappa/2\pi\sim1.5$~MHz$\sim0.0003g$ and using above parameters we can obtain a relatively high fidelity $F=0.993$. Therefore, our scheme is absolutely possible to be implemented with the current experimental technology.

\section{Conclusion}
\label{sec5}
In conclusion, we have proposed a scheme to generate three-dimensional entanglement between two atoms trapped in two cavities connected by a fiber via invariant-based shortcut to adiabatic passage. With the help of quantum Zeno dynamics, the invariant-based shortcut is constructed for the generation of two-atom three-dimensional entanglement. Based on our scheme, the operation time for generation of three-dimensional entanglement is short and not necessary to be precisely controlled. The numerical simulations show that our scheme is robust against the variations in the parameters and the decoherence induced by the atomic spontaneous emissions and the cavity-fiber photon leakages. In shorts, our scheme is robust, effective and fast. Moreover, the discussion on the feasibility indicates that our scheme is quite possible to be implemented with the current experimental technology.

\begin{center}
{\bf{ACKNOWLEDGMENT}}
\end{center}
This work was supported by the National Natural Science Foundation of China under Grant Nos. 11464046 and 61465013.

\end{document}